%Paper: hep-ph/9410298
%From: Hamzaoui <hamzaoui@mercure.phy.uqam.ca>
%Date: Sun, 16 Oct 1994 19:04:02 -0400

\magnification 1200
\input epsf.tex
\baselineskip=11pt\lineskiplimit=11pt\lineskip=11pt
%\voffset=
\line{\hfil UQAM-PHE/94-10}
\line{\hfil September 1994}
\vskip0.5cm
\centerline{\bf A Detailed Analysis of a Modified Fritzsch}
\centerline{\bf Scheme of Quark Mass Matrices}
\vskip2cm
\centerline{\bf E. Boridy, C. Hamzaoui\footnote*{
hamzaoui@mercure.phy.uqam.ca},
F. Lemay and J. Lindig}
\vskip1cm
\centerline{D\'epartement de Physique,
Universit\'e du Qu\'ebec \`a Montr\'eal,}
\centerline{Case Postale 8888,
Succ. Centre-Ville, Montr\'eal, Qu\'ebec, Canada, H3C 3P8.}
\vskip2cm
\centerline{\bf ABSTRACT}
\vskip1cm
We investigate in detail a modified version of the Fritzsch scheme where the
diagonal entries  $(M_{u,d})_{22}$ are introduced as free parameters instead
of being zero. We find that $(M_u)_{22}$ and $(M_d)_{22}$ are restricted to lie
in the range $m_u-m_c\leq (M_u)_{22}< m_t-m_c$ and
$m_d-m_s\leq (M_d)_{22}< m_b-m_s$. In the context of this scheme,
we obtain the ratio ${|V_{td}|\over |V_{ts}|}\approx\sqrt{{m_d\over m_s}}$
under quite general conditions whereas the ratio ${|V_{ub}|\over |V_{cb}|}$
is sensitive to the values of $(M_u)_{22}$ and $(M_d)_{22}$. The two
independent phases are found to take rather preferred values.
\vfill\eject
{\bf I. Introduction}\medskip

In an attempt to construct a theory in accordance with the observed quark mass
spectrum [1] and quark mixings [2], many speculations have been made
and some ansatzes were proposed [3]. A popular ansatz is the one
suggested by Fritzsch [4].
However, it is known that this scheme predicts a top quark mass less
than 90 GeV [5]. In view of the mounting evidence in favour of a heavy
top quark [6], $m_t=174\pm10$ GeV, the Fritzsch scheme is ruled out.\par
In this paper, we propose to study in details a modification of
the Fritzsch scheme of quark mass matrices by allowing some of their
elements to be nonzero. To be more specific, we take the up-type
and down-type quark mass matrices $M_u$ and $M_d$ to be of the
following generic form,
$$\eqalign{
M_u=\pmatrix{
0   & x   & 0 \cr
x^* & \alpha   & b \cr
0   & b^* & a \cr
},\phantom{pp}
M_d=\pmatrix{
0   & y   & 0 \cr
y^* & \beta   & f \cr
0   & f^* & d \cr}
.}\eqno(1)$$
In fact, such a modification has been considered in the literature [7].
However, the extra parameters added were dictated a certain value in an
ad-hoc manner. In our analysis, we keep these additional parameters
$\alpha$ and $\beta$ arbitrary and determine their parameter space
in general. We shall also discuss the issue of the phases and show
that under general circumstances the two independent phases of the ansatz
take some preferred values. We find that $\alpha$ and $\beta$ are restricted
to lie in the range $m_u-m_c\leq \alpha < m_t-m_c$ and
$m_d-m_s\leq \beta < m_b-m_s$. The obtained allowed range is further
restricted by comparison with experiments on the values of the elements
of the Kobayashi-Maskawa mixing matrix [8].\par
Several features emerge from this study. For instance, it turns out that the
mixing of the first two generations is very little affected by the presence of
the two parameters $\alpha$ and $\beta$. We show also that the third generation
decouples as $\alpha$ and $\beta$ approach their lower bounds.
At this point, it is interesting to note that the obtained mixing matrix
in this limit is equal to the one in the limit of infinite third generation
masses. Then we show that the ratio
${|V_{td}|\over |V_{ts}|}\approx\sqrt{{m_d\over m_s}}$
is obtained under quite general conditions. In contrast, the ratio
${|V_{ub}|\over |V_{cb}|}$
is very sensitive to the values of the parameters $\alpha$ and $\beta$.
In Sec. II, we calculate the mixing matrix elements as a function of the mass
matrices parameters and investigate the possible ranges which are allowed for
$\alpha$ and $\beta$. In Sec. III, we outline the special features possessed by
this
scheme and discuss our results. Our conclusions are contained in Sec. IV.
\bigskip
{\bf II. Structure of the modified scheme of mass matrices}
\medskip
In this section we will study the structure of the scheme represented by
the matrices of eq. (1). These matrices are hermitian. Consequently, the
parameters $a$, $d$, $\alpha$ and $\beta$ are restricted to be real whereas
$x$, $y$, $b$ and $f$ are in general complex numbers.
However, the pair $x$ and $b$ or $y$ and $f$ can be made real by performing a
suitable phase transformation $P$ on the quark fields. $P$ is given by
$$\eqalign{
P_{ij}=e^{i\phi_i}\delta_{ij}
.}\eqno(2)$$
The transformed mass matrices $P^{\dag}M_uP$ and $P^{\dag}M_dP$ become:
$$\eqalign{
P^{\dag}M_uP=\pmatrix{
0 &|x|e^{i(\delta_x+\phi_2-\phi_1)} & 0\cr
|x|e^{-i(\delta_x+\phi_2-\phi_1)}  & \alpha  &
|b|e^{i(\delta_b+\phi_3-\phi_2)}\cr
0 &|b|e^{-i(\delta_b+\phi_3-\phi_2)}  & a \cr}
}$$
$$\eqalign{
P^{\dag}M_dP=\pmatrix{
0 &|y|e^{i(\delta_y+\phi_2-\phi_1)}  & 0\cr
|y|e^{-i(\delta_y+\phi_2-\phi_1)}    & \beta  &
|f|e^{i(\delta_f+\phi_3-\phi_2)}\cr
0 &|f|e^{-i(\delta_f+\phi_3-\phi_2)} & d \cr}
.}\eqno(3)$$
Now we can use this freedom of rephasing and choose
$\delta_f+\phi_3-\phi_2=\delta_y+\phi_2-\phi_1=0$ such that
the elements $y$ and $f$ are real. In the following,
the two independent phases are always associated with the elements
$x=|x|e^{i\delta_x}$ and $b=|b|e^{i\delta_b}$.\par
At this stage, some properties of hermitian matrices
are worth mentioning. The diagonal entries of a hermitian matrix are bounded by
its smallest and largest eigenvalue respectively. Consequently, having diagonal
zero entries in a $3\times 3$ hermitian matrix is only possible if one of its
eigenvalues has the opposite sign of the other two.
There is no contradiction since the quarks are described by fermion fields
which
do not fix the sign of the masses. To be definite, in the following the
diagonal
form of the mass matrices is assumed to be
$$\eqalign{
M^D_u=\pmatrix{
m_u & 0    & 0   \cr
0   & -m_c & 0   \cr
0   & 0    & m_t \cr
},\phantom{pp}
M^D_d=\pmatrix{
m_d & 0    & 0   \cr
0   & -m_s & 0   \cr
0   & 0    & m_b \cr
}
.}\eqno(4)$$
It should be noted that beside the signs of the eigenvalues, this
convention fixes also the labeling of the axis in generation space.\par
In order to study the implications of this ansatz, we express the mixing
matrix elements in terms of the parameters of ansatz (1).
The characteristic equations of the mass matrices (1) are
$$\eqalignno{
0&=(a-\lambda )(\lambda(\lambda-\alpha )-|x|^2)+\lambda |b|^2 \cr
0&=(d-\gamma )(\gamma(\gamma-\beta)-|y|^2)+\gamma |f|^2
.&(5)}$$
Denoting the eigenvalues by $\lambda_i$ for the up-sector and $\gamma_j$
for the down-sector, one obtains the eigenvectors of $M_u$ and $M_d$:
$$\eqalign{
e_{ui}={1 \over N_i} \left(1, {\lambda_i \over x}
, {\lambda_i b^* \over x(\lambda_i-a)}\right),
\phantom{pp}
e_{dj}={1 \over M_j} \left(1, {\gamma_j \over y}
, {\gamma_j f^* \over y(\lambda_j-d)}\right)
}\eqno(6)$$
$N_i$ and $M_j$ being the normalization constants:
$$\eqalign{
N^2_i=1+{\lambda_i^2 \over |x|^2}\left(
1+{|b|^2 \over (\lambda_i-a)^2}\right),
\phantom{pp}
M^2_j=1+{\gamma_j^2 \over |y|^2}\left(
1+{|f|^2 \over (\gamma_j-d)^2}\right)
.}\eqno(7)$$
The knowlegde of the eigenvectors of $M_u$ and $M_d$ allows one
immediately to write down the unitary matrices $U_u$ and $U_d$ which
diagonalize $M_u$ and $M_d$:
$$\eqalignno{
U_u&=\pmatrix{
{1 \over N_1}   & {1 \over N_2}   & {1 \over N_3} \cr
{\lambda_1 \over N_1x} & {\lambda_2 \over N_2x}   & {\lambda_3 \over N_3x} \cr
  {\lambda_1b^* \over N_1x(\lambda_1 -a)}
& {\lambda_2b^* \over N_2x(\lambda_2 -a)}
& {\lambda_3b^* \over N_3x(\lambda_3 -a)}
}\cr
U_d&=\pmatrix{
{1 \over M_1}   & {1 \over N_2}   & {1 \over M_3} \cr
{\gamma_1 \over M_1y} & {\gamma_2 \over M_2y}   & {\gamma_3 \over M_3y} \cr
  {\gamma_1f^* \over M_1y(\gamma_1 -d)}
& {\gamma_2f^* \over M_2y(\gamma_2 -d)}
& {\gamma_3f^* \over M_3y(\gamma_3 -d)} \cr
}
.&(8)}$$
The mixing matrix is defined as $U^{\dag}_uU_d$. The mixing matrix
elements are then given by:
$$\eqalign{
V_{ij}={1 \over N_iM_j}\left(1+{\lambda_i\gamma_j \over x^*y}
\left(1+{bf^* \over (\lambda_i-a)(\gamma_j-d)}\right)\right)
.}\eqno(9)$$
Expressing the parameters of ansatz (1) in terms of
the quark masses by using the six invariants $Tr(M_i)$,
$Tr(M^2_i)$ and $Det(M_i)$ $(i=u,d)$, one finds:
$$\eqalignno{
Tr(M_u)&=a+\alpha=m_u-m_c+m_t,\phantom{pp}
Tr(M_d)=d+\beta=m_d-m_s+m_b \cr
Tr(M^2_u)&=2(|x|^2+|b|^2)+a^2+\alpha^2=m_u^2+m_c^2+m_t^2 ,\cr
Tr(M^2_d)&=2(|y|^2+|f|^2)+d^2+\beta^2=m_d^2+m_s^2+m_b^2 ,\cr
Det(M_u)&=-a|x|^2=-m_um_cm_t,\phantom{pp}
Det(M_d)=-d|y|^2=-m_dm_sm_b
.&(10)}$$
When solved for $x$, $y$, $b$ and $f$, these relations yield:
$$\eqalignno{
a&=m_u-m_c+m_t-\alpha,\phantom{pp} d=m_d-m_s+m_b-\beta\cr
|x|^2&={ m_um_cm_t \over m_u-m_c+m_t-\alpha} ,\phantom{pp}
|y|^2={ m_dm_sm_b \over m_d-m_s+m_b-\beta} \cr
|b|^2&={ (m_c-m_u+\alpha)(m_t-m_c-\alpha)(m_t+m_u-\alpha)
\over m_u-m_c+m_t-\alpha} \cr
|f|^2&={ (m_s-m_d+\beta)(m_b-m_s-\beta)(m_b+m_d-\beta)
\over m_d-m_s+m_b-\beta}
.&(11)}$$
The mixing matrix is now given in terms of the quark masses, the two
parameters $\alpha$, $\beta$ and the phases $\delta_x$ and $\delta_b$
by substituting eq. (11) in eq. (9). At this point,
it is interesting to note that in the case of the Fritzsch ansatz
$\alpha=\beta=0$, all parameters are entirely expressed in terms
of the masses and phases alone. This is exactly what makes it
possible to approximate the mixing matrix only knowing the magnitude of the
quark mass ratios. In that sense, the situation has now tremendously changed
because the quality of any approximation will crucially depend on
the magnitude of the parameters $\alpha$ and $\beta$ which are so
far unknown. It is therefore important to get some information on the
values of $\alpha$ and $\beta$.\par
As mentioned earlier, the diagonal entries of any hermitian matrix
are bounded by its largest and smallest eigenvalues. According to
the sign convention (4) and bearing in mind that $m_u<m_c<m_t$
and $m_d<m_s<m_b$, this fact subscribes the values of
$\alpha$ and $\beta$ to the range $-m_c\leq \alpha \leq m_t$ and
$-m_s\leq \beta \leq m_b$ respectively.
Due to the specific form of the ansatz (1), those intervals
can be further narrowed.
The positivity of $|x|^2$
immediately yields the condition $\alpha<m_u-m_c+m_t$. Furthermore,
$|b|^2\geq 0$ is only satisfied if the factor
$(m_c-m_u+\alpha)(m_t-m_c-\alpha)$ is non-negative. This is guaranteed
if $m_u-m_c\leq \alpha \leq m_t-m_c$. If $\alpha$ equals one of these
bounds, then $|b|$ equals zero. In this case, the parameter $a$ becomes
directly an eigenvalue of $M_u$ as can be seen from (1). On the
other hand, $a$ has to obey the condition $a=m_u-m_c+m_t-\alpha$.
So $\alpha=m_u-m_c$ gives correctly $a=m_t$, while $\alpha=m_t-m_c$ results
in $a=m_u$ which contradicts the convention (4) for the labeling of the axes.
The allowed range for the values of $\beta$ can be found in an analogous
manner.
Finally, one obtains:
$$\eqalign{
m_u-m_c\leq \alpha < m_t-m_c ,\phantom{ppp}
m_d-m_s\leq \beta < m_b-m_s
.}\eqno(12)$$
\vfill\vfill\eject
{\bf III. Results and Discussions}\medskip
Before we analyze the exact expressions of the Kobayashi-Maskawa
mixing matrix, we first study its behaviour in the limits
$\alpha \to m_u-m_c$, $\beta \to m_d-m_s$ and $m_t, m_b \to \infty$.
As shown in Sec. II, the parameters $|b|$ and $|f|$
vanish in the limit $\alpha \to m_u-m_c$ and $\beta \to m_d-m_s$.
Hence, the third generation decouples or, in other words, the mass matrices
have a simultaneous eigenvector which results in a real mixing matrix.
In the limit $\alpha \to m_u-m_c$, one finds for the following parameters:
$$\eqalignno{
&\lim_{\alpha \to m_u-m_c}N_u^{-2}={m_c \over m_u+m_c},\phantom{p}
\lim_{\alpha \to m_u-m_c}N_c^{-2}={m_u \over m_u+m_c},\phantom{p}
\lim_{\alpha \to m_u-m_c}N_t^{-2}= 0 \cr
&\lim_{\alpha \to m_u-m_c}{m_u^2 \over N_u^2|x|^2}={m_u \over m_u+m_c},
\phantom{p}
\lim_{\alpha \to m_u-m_c}{m_c^2 \over N_c^2|x|^2}={m_c \over m_u+m_c}, \cr
&\lim_{\alpha \to m_u-m_c}{m_t^2 \over N_t^2|x|^2}=0,
\phantom{p}
\lim_{\alpha \to m_u-m_c}{m_u^2|b|^2 \over N_u^2|x|^2(m_t-m_c)^2}=0, \cr
&\lim_{\alpha \to m_u-m_c}{m_c^2|b|^2 \over N_c^2|x|^2(m_t+m_u)^2}=0,
\phantom{p}
\lim_{\alpha \to m_u-m_c}{m_t^2|b|^2 \over N_c^2|x|^2(m_c-m_u)^2}=1
.&(13)}$$
A set of similar relations holds for the parameters of the down-type mass
matrix.
Hence, the mixing matrix becomes:
$$\eqalignno{
V_{ud}&=\sqrt{{m_cm_s\over (m_u+m_c)(m_d+m_s)}}
+\sqrt{{m_um_d\over (m_u+m_c)(m_d+m_s)}}e^{i\delta_x} \cr
V_{us}&=\sqrt{{m_cm_d\over (m_u+m_c)(m_d+m_s)}}
-\sqrt{{m_um_s\over (m_u+m_c)(m_d+m_s)}}e^{i\delta_x} \cr
V_{cd}&=\sqrt{{m_um_s\over (m_u+m_c)(m_d+m_s)}}
-\sqrt{{m_cm_d\over (m_u+m_c)(m_d+m_s)}}e^{i\delta_x} \cr
V_{cs}&=\sqrt{{m_um_d\over (m_u+m_c)(m_d+m_s)}}
+\sqrt{{m_cm_s\over (m_u+m_c)(m_d+m_s)}}e^{i\delta_x} \cr
V_{ub}&=V_{cb}=V_{td}=V_{ts}=0, \phantom{pp}
V_{tb}=e^{i(\delta_x+\delta_b)}
&(14)}$$
In this case, the third generation decouples and the phase $\delta_b$
disappears. Using the triangle inequality, one then finds the following
estimation for the element $|V_{us}|$:
$$\eqalign{
\left| {\sqrt{m_cm_d}-\sqrt{m_um_s}\over \sqrt{(m_u+m_c)(m_d+m_s)}} \right|
\leq |V_{us}| \leq {\sqrt{m_cm_d}+\sqrt{m_um_s}\over \sqrt{(m_u+m_c)(m_d+m_s)}}
}\eqno(15)$$
which agrees rather well with experiment. From the expression for $|V_{us}|$,
we obtain for the phase $\delta_x$:
$$\eqalign{
\cos\delta_x ={m_cm_d+m_um_s-(m_u+m_c)(m_d+m_s)|V_{us}|^2
\over \sqrt{4m_um_cm_dm_s}}
.}\eqno(16)$$
This relation gives a value of approximately $\pi /2$ for the phase $\delta_x$.
Taking into account the actual scale of the quark masses, one may approximate
the relations (15) and (16) by
$$\eqalign{
\left| \sqrt{{m_d \over m_s}}-\sqrt{{m_u \over m_c}} \right| \leq |V_{us}|
\leq \sqrt{{m_d \over m_s}}+\sqrt{{m_u \over m_c}}
}\eqno(17)$$
and
$$\eqalign{
\cos\delta_x =\left( {m_s \over m_b}+{m_u \over m_c}-|V_{us}|^2
\right)
\left(4 {m_dm_u \over m_sm_c}
\right)^{-{1 \over 2}}
.}\eqno(18)$$
In this limit, the mixing between the first two generations is correctly
described. Eq. (15) yields appropriate bounds for the Cabbibo angle.
The decoupling of the third generation happens regardless of the scale
of the top and bottom quark masses as the parameters $\alpha$ and $\beta$
approach their lower bounds. This motivates us to investigate
the following expansion. Introducing the expansion parameters
$\varepsilon_1={\alpha+m_c-m_u\over m_t}$ and
$\varepsilon_2={\beta+m_s-m_d\over m_b}$ and
displaying only terms up to first order, one obtains for the up-type
parameters:
$$\eqalignno{
N_u^{-2}&={m_c \over m_u+m_c}
\left(1-\varepsilon_1{m_u \over m_t-m_u}\right)+h.o.t.,\cr
N_c^{-2}&={m_u \over m_u+m_c}
\left(1+\varepsilon_1{m_c \over m_t+m_c}\right)+h.o.t.,\cr
N_t^{-2}&={\varepsilon_1m_um_c \over (m_t+m_c)(m_t-m_u)}+h.o.t,\cr
{m_u^2 \over N_u^2|x|^2}&={m_u \over m_u+m_c}
\left(1-\varepsilon_1{m_t\over m_t-m_u}\right)+h.o.t.,\cr
{m_c^2 \over N_c^2|x|^2}&={m_c \over m_u+m_c}
\left(1-\varepsilon_1{m_t\over m_t+m_c}\right)+h.o.t.,\cr
{m_t^2 \over N_t^2|x|^2}&=
{\varepsilon_1m_t^2 \over (m_t+m_c)(m_t-m_u) }+h.o.t.,\cr
{m_u^2|b|^2 \over N_u^2|x|^2(m_t-m_c)^2}&=
\varepsilon_1{m_u(m_t+m_c) \over (m_u+m_c)(m_t-m_u)}+h.o.t.,\cr
{m_c^2|b|^2 \over N_c^2|x|^2(m_t+m_u)^2}&=
\varepsilon_1{m_c(m_t-m_u) \over (m_u+m_c)(m_t+m_c)}+h.o.t.,\cr
{m_t^2|b|^2 \over N_c^2|x|^2(m_c-m_u)^2}&=
1-\varepsilon_1{(m_t^2+m_um_c) \over (m_t+m_c)(m_t-m_u)}+h.o.t.
&(19)}$$
where h.o.t. stands for higher order terms.
Similar relations hold for the down-type parameters.
Up to this order, the off-diagonal elements of the mixing matrix become:
$$\eqalignno{
V_{us}=&\sqrt{ {m_c \over (m_u+m_c) } {m_d \over (m_d+m_s) }
\left(1-\varepsilon_1{m_u\over m_t-m_u}\right)
\left(1+\varepsilon_2{m_s\over m_b+m_s}\right)} \cr
-&\sqrt{ {m_u \over (m_u+m_c) }{m_s \over (m_d+m_s) }
\left(1-\varepsilon_1{m_t\over m_t-m_u}\right)
\left(1-\varepsilon_2{m_b\over m_b+m_s}\right)}
e^{i\delta_x} \cr
-&\sqrt{\varepsilon_1\varepsilon_2
{m_um_s(m_t+m_c)(m_b-m_d) \over (m_t-m_u)(m_u+m_c)(m_b+m_s)(m_d+m_s)}
}e^{i(\delta_x+\delta_b)}
,&(20)}$$
$$\eqalignno{
V_{ub}=&\sqrt{ {m_c \over (m_u+m_c)}
\left(1-\varepsilon_1{m_u\over m_t-m_u}\right)
{\varepsilon_2m_dm_s \over (m_b+m_s)(m_b-m_d)}}\cr
+&\sqrt{{m_u \over (m_u+m_c) }\left(1-\varepsilon_1{m_t\over m_t-m_u}\right)
{\varepsilon_2m_b^2 \over (m_b+m_s)(m_b-m_d)}
}e^{i\delta_x}\cr
-&\sqrt{ \varepsilon_1{m_u(m_t+m_c)\over (m_t-m_u)(m_u+m_c)}
\left(1-{\varepsilon_2(m_b^2+m_dm_s)\over (m_b+m_s)(m_b-m_d)}\right)
}e^{i(\delta_x+\delta_b)}
,&(21)}$$
and
$$\eqalignno{
V_{cb}=&\sqrt{ {m_u \over (m_u+m_c) }
\left(1+\varepsilon_1{m_c\over m_t+m_c}\right)
{\varepsilon_2m_dm_s \over (m_b+m_s)(m_b-m_d)}
}\cr
-&\sqrt{{m_c \over (m_u+m_c) }\left(1-\varepsilon_1{m_t\over m_t+m_c}\right)
{\varepsilon_2m_b^2 \over (m_b+m_s)(m_b-m_d)}}
e^{i\delta_x}\cr
+&\sqrt{\varepsilon_1
{m_c(m_t-m_u)\over (m_t+m_c)(m_u+m_c)}
\left(1-{\varepsilon_2(m_b^2+m_dm_s)\over (m_b+m_s)(m_b-m_d)}\right)
}e^{i(\delta_x+\delta_b)}
.&(22)}$$
The lower off-diagonal elements are found by replacing all up-type
by the corresponding down-type indices. It should be mentioned
that from the bounds on $\alpha$ and $\beta$ given in eq. (12),
the parameters $\varepsilon_1$ and $\varepsilon_2$ vary in the following
range:
$$\eqalign{
&0\le\varepsilon_1< {m_t-m_u \over m_t}\approx 1\cr
&0\le\varepsilon_2< {m_b-m_d \over m_b}\approx 1\cr
}\eqno(23)$$
This expansion will be relevant when we deal with the approximation later.
Injecting the values $\varepsilon_1=\varepsilon_2=0$ clearly reproduces eq.
(14).
Another interesting case is the limit $m_t,m_b \to \infty$.
To study the behaviour of the mixing matrix in this limit,
we expand its elements in terms of the ratios $m_{u,c}/m_t$, $\alpha /m_t$,
$m_{d,s}/m_b$ and $\beta /m_b$. Keeping only the linear terms in those
ratios, one finds:
$$\eqalignno{
N_u^{-2}&\approx{m_c \over m_u+m_c},\phantom{p}
N_c^{-2}\approx{m_u \over m_u+m_c},\phantom{p}
N_t^{-2}\approx 0 \cr
{m_u^2 \over N_u^2|x|^2}&\approx{m_u \over m_u+m_c}
\left(1-{m_c-m_u+\alpha \over m_t} \right), \cr
{m_c^2 \over N_c^2|x|^2}&\approx{m_c \over m_u+m_c}
\left(1-{m_c-m_u+\alpha \over m_t} \right), \cr
{m_t^2 \over N_t^2|x|^2}&\approx{m_c-m_u+\alpha \over m_t} \cr
{m_u^2|b|^2 \over N_u^2|x|^2(m_t-m_c)^2}&\approx
{m_u \over m_u+m_c}{m_c-m_u+\alpha \over m_t},\cr
{m_c^2|b|^2 \over N_c^2|x|^2(m_t+m_u)^2}&\approx
{m_c \over m_u+m_c}{m_c-m_u+\alpha \over m_t},\cr
{m_t^2|b|^2 \over N_c^2|x|^2(m_c-m_u)^2}&\approx
1-{m_c-m_u+\alpha \over m_t}-{\alpha \over m_t}
.&(24)}$$
Similar relations hold for the down-type parameters.
For simplicity, we introduce the following notation:
$$\eqalignno{
\sin^2\theta_1&={m_u \over m_u+m_c},\phantom{pp}
\sin^2\theta_2={m_d \over m_d+m_s}=\varepsilon_1 \cr
\sin^2\phi_1&={m_c-m_u+\alpha \over m_t},\phantom{pp}
\sin^2\phi_2={m_s-m_d+\beta \over m_b}=\varepsilon_2
.&(25)}$$
The mixing matrix elements then become:
$$\eqalignno{
V_{ud}=&\cos\theta_1\cos\theta_2+\sin\theta_1\cos\phi_1\sin\theta_2\cos\phi_2
e^{i\delta_x}\cr
&+\sin\theta_1\sin\phi_1\sin\theta_2\sin\phi_2
e^{i(\delta_x+\delta_b)}\cr
V_{us}=&\cos\theta_1\sin\theta_2-\sin\theta_1\cos\phi_1\cos\theta_2\cos\phi_2
e^{i\delta_x}\cr
&-\sin\theta_1\sin\phi_1\cos\theta_2\sin\phi_2
e^{i(\delta_x+\delta_b)}\cr
V_{ub}=&\sin\theta_1\cos\phi_1\sin\phi_2e^{i\delta_x}
-\sin\theta_1\sin\phi_1\left(\cos\phi_2-{\beta \over m_b}\right)
e^{i(\delta_x+\delta_b)}\cr
V_{cd}=&\sin\theta_1\cos\theta_2-\cos\theta_1\cos\phi_1\sin\theta_2\cos\phi_2
e^{i\delta_x}\cr
&-\cos\theta_1\sin\phi_1\sin\theta_2\sin\phi_2
e^{i(\delta_x+\delta_b)}\cr
V_{cs}=&\sin\theta_1\sin\theta_2+\cos\theta_1\cos\phi_1\cos\theta_2\cos\phi_2
e^{i\delta_x}\cr
&+\cos\theta_1\sin\phi_1\cos\theta_2\sin\phi_2
e^{i(\delta_x+\delta_b)}\cr
V_{cb}=&-\cos\theta_1\cos\phi_1\sin\phi_2e^{i\delta_x}+
\cos\theta_1\sin\phi_1\left(\cos\phi_2-{\beta \over m_b}\right)
e^{i(\delta_x+\delta_b)}\cr
V_{td}=&\sin\phi_1\sin\theta_2\cos\phi_2e^{i\delta_x}-
\left(\cos\phi_1-{\alpha \over m_t}\right)\sin\theta_2\sin\phi_2
e^{i(\delta_x+\delta_b)}\cr
V_{ts}=&-\sin\phi_1\cos\theta_2\cos\phi_2e^{i\delta_x}+
\left(\cos\phi_1-{\alpha \over m_t}\right)\cos\theta_2\sin\phi_2
e^{i(\delta_x+\delta_b)}\cr
V_{tb}=&\sin\phi_1\sin\phi_2e^{i\delta_x}
+\left(\cos\phi_1-{\alpha \over m_t}\right)
\left(\cos\phi_2-{\beta \over m_b}\right)
e^{i(\delta_x+\delta_b)}
.&(26)}$$
Performing the limit $m_t,m_b \to \infty$ produces the mixing matrix
elements
$$\eqalignno{
V_{ud}&=\cos\theta_1\cos\theta_2+\sin\theta_1\sin\theta_2e^{i\delta_x} \cr
V_{us}&=\cos\theta_1\sin\theta_2-\cos\theta_1\cos\theta_2e^{i\delta_x} \cr
V_{cd}&=\sin\theta_1\cos\theta_2-\cos\theta_1\sin\theta_2e^{i\delta_x} \cr
V_{cs}&=\sin\theta_1\sin\theta_2+\cos\theta_1\cos\theta_2e^{i\delta_x} \cr
V_{ub}&=V_{cb}=V_{td}=V_{ts}=0, \phantom{pp}
V_{tb}=e^{i(\delta_x+\delta_b)}
&(27)}$$
There are several remarks worth mentioning about the above results.
The mixing matrix elements given in eq. (27) are exactly the same as
 the ones obtained
in the limit $\alpha \to m_u-m_c$ and $\beta \to m_d-m_s$.
This is expected since in both cases the third generation decouples.
Here again, the value of $|V_{us}|$ is correctly bounded and the phase
$\delta_x$ is approximately $\pi /2$.\par
The ratios $|V_{ub}|/|V_{cb}|$ and
$|V_{td}|/|V_{ts}|$ can now be calculated from equation (26):
$$\eqalign{
{|V_{ub}|\over |V_{cb}|}={\sin\theta_1 \over \cos\theta_1}
=\sqrt{{m_u \over m_c}}
\phantom{ppp}
{\rm and}
\phantom{ppp}
{|V_{td}|\over |V_{ts}|}={\sin\theta_2 \over \cos\theta_2}
=\sqrt{{m_d\over m_s}}
.}\eqno(28)$$
Again, these are the ratios which were obtained by the method
of the expansion in the parameters $\varepsilon_1={\alpha+m_c-m_u\over m_t}$
and
$\varepsilon_2={\beta+m_s-m_d\over m_b}$.\par
The considered limits certainly cannot describe the real world
where the third generation mixings are non-zero.
Furthermore, CP-violation is absent in these limits. However, the expansions
up to the first order in the respective parameters have exhibited some
 interesting
properties such as the values of the ratios $|V_{ub}|/|V_{cb}|$ and
$|V_{td}|/|V_{ts}|$.\par
Now we analyze the parameter space spanned by $\alpha$, $\beta$
and the phases from the exact expressions of the $|V_{ij}|$ given in
Sec. II. This will serve to probe their order of magnitude.
Therefore, it is sufficient to use the central values for the quark
masses as input in all plots; $m_u(1GeV)=5$ MeV, $m_d(1GeV)=8$ MeV,
 $m_c(1GeV)=1.35$ GeV, $m_s(1GeV)=170$ MeV,
$m_b(1GeV)=5.5$ GeV and $m_t(1GeV)=280$ GeV.
The value of $m_t$ at 1 GeV corresponds to the central value of the
physical mass of
174 GeV [6]. \par
Using the triangle inequality, one is able to estimate the
bounds of the magnitudes $|V_{ij}|$.
In particular, there is one interesting lower bound which does not depend
on the phase $\delta_x$ and which is therefore suitable
 for a graphical analysis.
It is given by
$$\eqalign{
|V_{ij}|\geq \left|{1 \over N_iM_j}-
{|\lambda_i||\gamma_j| \over N_iM_j|x||y|}
\left|1+{|b||f|e^{i\delta_b}  \over (\lambda_i-a)(\gamma_j-d)} \right|\right|
.}\eqno(29)$$
We plot the lower bound on $|V_{cb}|$ from eq. (29) versus $\alpha$ and $\beta$
for representative values of the phase $\delta_b$, namely $0$, ${\pi\over 2}$
and $\pi$.
These plots are displayed in Figs. 1 to 4. Amazingly enough,
one can see from these figures that the solutions corresponding to the
 cases ${\pi\over 2}$ and $\pi$
are included in the case $\delta_b=0$. This just reflects the decoupling of the
third generation in the limit $\alpha \to m_u-m_c$ and $\beta \to m_d-m_s$
in which the phase $\delta_b$ is undetermined. As has already been mentioned,
this happens regardless of the value of the phase $\delta_x$. However, we know
from eq. (16) that the phase $\delta_x$ is approximately ${\pi \over 2}$.
Without loss of generality, we investigate the parameter space of $\alpha$,
 $\beta$
for values of $\delta_x$ around ${\pi\over 2}$ and for $\delta_b$ around zero.
This is illustrated in Fig. 5 where we plot the exact expression for $|V_{cb}|$
versus $\alpha$ and $\beta$ using these values for the phases.
Fig. 6 shows the lower bound on $|V_{ub}|$ which corresponds to eq. (29) for
$\delta_b=0$. In Fig. 7, we display the exact expression for $|V_{ub}|$
versus $\alpha$ and $\beta$ for $\delta_x={\pi \over 2}$
and $\delta_b=0$. This suggests that the parameters $\alpha$ and $\beta$
are much
smaller than $m_t$ and $m_b$ respectively and therefore that
 $\varepsilon_1\ll 1$
and $\varepsilon_2\ll 1$. For completeness, the
exact expression for $|V_{us}|$ is shown in Fig. 8 as a function of
 $\alpha$ and $\beta$ for
$\delta_x={\pi \over 2}$ and $\delta_b=0$ to show that it is compatible
with this range for the parameters $\alpha$ and $\beta$.\par
The above results allow us to describe the mixing matrix elements by the
 first order
expansion previously advocated in eqs. (20) to (22) in terms of
$\varepsilon_1$ and $\varepsilon_2$. Taking into account the observed
hierarchy of masses , we obtain for the mixing matrix elements:
$$\eqalignno{
V_{us}\approx&\sqrt{{m_d\over m_s}}-\sqrt{{m_u\over m_c}}e^{i\delta_x},\cr
V_{ub}\approx&\sqrt{\varepsilon_2{m_dm_s \over m_b^2}}
+\sqrt{m_u \over m_c}\left({\sqrt{ \varepsilon_2}
-e^{i\delta_b}\sqrt{\varepsilon_1}
}\right)e^{i\delta_x}\cr
V_{cb}\approx&-\left({\sqrt{\varepsilon_2}-e^{i\delta_b}\sqrt{\varepsilon_1}
}\right)e^{i\delta_x}
.&(30)}$$
The expressions for $V_{cd}$, $V_{td}$ and $V_{ts}$ are obtained
by just interchanging corresponding up-type and down-type
indices as well as $\varepsilon_1$ and $\varepsilon_2$ in eq. (30).
Note that $V_{ub}$ and $V_{td}$ can be written as
$$\eqalign{
V_{ub}\approx\sqrt{\varepsilon_2{m_dm_s \over m_b^2}}-\sqrt{{m_u\over
m_c}}V_{cb}
}$$
and
$$\eqalign{
V_{td}\approx\sqrt{\varepsilon_1{m_um_c \over m_t^2}}-\sqrt{{m_d\over
m_s}}V_{ts}
.}\eqno(31)$$
Using the triangle inequality, we obtain the following bounds on the ratios
${|V_{ub}|\over |V_{cb}|}$ and ${|V_{td}|\over |V_{ts}|}$
$$\eqalign{
\left| \sqrt{\varepsilon_2{m_dm_s\over m_b^2|V_{cb}|^2}}-\sqrt{{m_u \over
m_c}}\right|\leq
{|V_{ub}|\over |V_{cb}| }\leq
\left| \sqrt{\varepsilon_2{m_dm_s\over m_b^2|V_{cb}|^2}}+\sqrt{{m_u \over
m_c}}\right|
.}\eqno(32)$$
$$\eqalign{
\left| \sqrt{\varepsilon_1{m_um_c\over m_t^2|V_{ts}|^2}}-\sqrt{{m_d \over
m_s}}\right|\leq
{|V_{td}|\over |V_{ts}| }\leq
\left| \sqrt{\varepsilon_1{m_um_c\over m_t^2|V_{ts}|^2}}+\sqrt{{m_d \over
m_s}}\right|
.}\eqno(33)$$
Some remarks are in order:\smallskip
i) The $\varepsilon_2$ dependent term in the expression of the ratio
 ${|V_{ub}|\over |V_{cb}|}$
is not always small in the allowed range of $\varepsilon_2$ and
therefore cannot
be neglegted. Indeed, this term could easily be of the order of the second
term $\sqrt{{m_u \over m_c}}$. This is in contrast to previous analyses [7]
where this term is absent.\smallskip
ii) In the case of the ratio ${|V_{td}|\over |V_{ts}|}$, the $\varepsilon_1$
dependent term
is always small compared to $\sqrt{{m_d \over m_s}}$ regardless of the value
of $\varepsilon_1$ ($0\leq \varepsilon_1 < 1$). Therefore, the ratio
${|V_{td}|\over |V_{ts}|}$ is very well approximated by:
$$\eqalign{
{|V_{td}|\over |V_{ts}|}\approx\sqrt{{m_d\over m_s}}
.}\eqno(34)$$\indent
iii) In view of eq. (30) and Fig. 8, the coupling of the first two
generations is only little affected by the coupling of the third one.
The expansion parameters $\varepsilon_1$ and $\varepsilon_2$ introduced
earlier, describe the effect of the third generation.\smallskip
iv) From Figs. 4, 5, 6 and 7, one can see that $\varepsilon_1\approx
\varepsilon_2$.
Injecting this in (30), one obtains for $|V_{cb}|$:
$$\eqalign{
|V_{cb}|\approx\sqrt{2\varepsilon_1(1-\cos\delta_b)}
.}\eqno(35)$$
As expected, as $\varepsilon_1$ tends to zero, $|V_{cb}|$ vanishes
regardless of the value of $\delta_b$.  This expression also
suggests that the phase $\delta_b$ is small in order to compensate the
increase in $\varepsilon_1$.\par
The relation $\varepsilon_1\approx \varepsilon_2$ leads directly to the
failure of the Fritzsch ansatz because with $\alpha=\beta=0$,
the relation $\varepsilon_1\approx \varepsilon_2$ implies
${m_c\over m_t}\approx{m_s\over m_b}$ which clearly violates the observed
pattern.\smallskip
v) From $|V_{cb}|$, we obtain an upper bound on $m_t$ as
$$\eqalign{
m_t\leq { m_c-m_u+\alpha \over
\left(|V_{cb}|-\sqrt{ {m_s-m_d+\beta \over m_b}}\right)^2}
.}\eqno(36)$$\bigskip
{\bf IV. Conclusions}\medskip
We have emphasized in our analysis the importance of the presence of
the parameters $\alpha$ and $\beta$ and their effect on the mixing matrix
elements. We have determined the possible range for these parameters
and found that $m_u-m_c\leq \alpha< m_t-m_c$ and $m_d-m_s\leq \beta< m_b-m_s$.
Based on the graphical analysis, we have found that the parameters
$\varepsilon_1$ and $\varepsilon_2$ are much smaller than one. Then, the
phases are found to take some preferred values, namely $\delta_b$ around
zero and $\delta_x$ around ${\pi\over 2}$.
This makes the first order expansion of the mixing matrix
elements in terms of the parameters $\varepsilon_1$ and $\varepsilon_2$
a suitable description of the Kobayashi-Maskawa matrix. Moreover,
we have shown that the ratio ${|V_{td}|\over |V_{ts}|}$ is independent of
$\varepsilon_1$ and $\varepsilon_2$ and is given
to a good
approximation by ${|V_{td}|\over |V_{ts}|}=\sqrt{{m_d\over m_s}}$.
In contrast,
the ratio ${|V_{ub}|\over |V_{cb}|}$ is very sensitive to the values
of $\varepsilon_1$ and $\varepsilon_2$.
\vskip2cm
{\bf Acknowledgements}\medskip
   We would like to thank F. Boudjema, G. Couture and S. Rajpoot for useful
discussions.

This research was partially funded by N.S.E.R.C. of Canada.

\vfill\vfill\eject
{\baselineskip=4pt\lineskiplimit=4pt\lineskip=4pt
{\bf Figure Captions}
\vskip2cm
Fig. 1: Plot of the lower bound on $|V_{cb}|$ as given by
eq. (29) in the range 0.035 to 0.055 versus $\alpha$ and $\beta$ for
$\delta_b=0$. Here and in all graphs, $\alpha$ and $\beta$ vary
in their allowed ranges.

\vskip1cm
Fig. 2: The same as in
Fig. 1 with $\delta_b={\pi \over 2}$.

\vskip1cm
Fig. 3: The same as in Fig. 1 with
$\delta_b=\pi$.

\vskip1cm
Fig. 4: Contour plot of $|V_{cb}|$ vs. $\alpha$ and $\beta$ for
 $\delta_b=0$.
The range $0.04\leq |V_{cb}| \leq 0.05$
is displayed.

\vskip1cm
Fig. 5: Contour plot of $|V_{cb}|$ vs. $\alpha$ and $\beta$ for
$\delta_b=0$ and $\delta_x={\pi \over 2}$.
The plot range is $0.04\leq |V_{cb}| \leq 0.05$.

\vskip1cm
Fig. 6: The lower bound, $|V_{ub}|=\left|{1 \over N_uM_b}-
{|m_u||m_b| \over N_uM_b|x||y|}
\left|1+{|b||f|e^{i\delta_b}  \over (m_u-a)(m_b-d)}
\right|\right|$, on $|V_{ub}|$ shown as a contour plot
 vs. $\alpha$ and $\beta$
for $\delta_b=0$ in the range $0.001\leq |V_{ub}| \leq 0.005$.

\vskip1cm
Fig. 7: Contour plot of $|V_{ub}|$ vs. $\alpha$ and $\beta$ for
$\delta_b=0$ and $\delta_x={\pi \over 2}$, where
the range is again  $0.001\leq |V_{ub}| \leq 0.005$.

\vskip1cm
Fig. 8: Contour plot of $|V_{us}|$ vs. $\alpha$ and $\beta$.
The phases are assigned the values
$\delta_b=0$ and $\delta_x={\pi \over 2}$ and $|V_{us}|$ is made to vary
in the range $0.15\leq |V_{us}| \leq 0.3$.
\vfill\vfill\eject
$$\vbox{\settabs
\+&ddddddddddddddddddddddddddddddddddddddddddddddddddddddddddddddd&\cr
\+&{\rm Fig. 1}\hfill& \cr
\+&\phantom{ppp}{\rm A lower bound on} $|V_{cb}|$ {\rm vs.} $\alpha$ {\rm and }
$\beta$ {\rm for} $\delta_b=0$\hfill& \cr
\+&\hfill\hfill& \cr
}$$
\epsfysize=7cm
\centerline{\epsffile[72 230 540 560]{/mercure/hamzaoui/pvcb1.eps}}
\vskip 0.1cm
$$\vbox{\settabs
\+&ddddddddddddddddddddddddddddddddddddddddddddddddddddddddddddddd&\cr
\+&{\rm Fig. 2}\hfill& \cr
\+&\phantom{ppp}{\rm A lower bound on} $|V_{cb}|$ {\rm vs.} $\alpha$
{\rm and } $\beta$ {\rm for} $\delta_b={\pi \over 2}$\hfill &\cr
\+&\hfill\hfill& \cr
}$$
\epsfysize=7cm
\centerline{\epsffile[72 230 540 560]{/mercure/hamzaoui/pvcb2.eps}}
\vfill\vfill\eject
$$\vbox{\settabs
\+&ddddddddddddddddddddddddddddddddddddddddddddddddddddddddddddddd&\cr
\+&{\rm Fig. 3}\hfill& \cr
\+&\phantom{ppp}{\rm A lower bound on} $|V_{cb}|$ {\rm vs.} $\alpha$
{\rm and } $\beta$ {\rm for} $\delta_b=\pi$\hfill &\cr
\+&\hfill\hfill&\cr
}$$
\epsfysize=7cm
\centerline{\epsffile[72 230 540 560]{/mercure/hamzaoui/pvcb3.eps}}
\vskip 0.5cm
$$\vbox{\settabs
\+&ddddddddddddddddddddddddddddddddddddddddddddddddddddddddddddddd&\cr
\+&{\rm Fig. 4}\hfill& \cr
\+&\phantom{ppp}{\rm A lower bound on} $|V_{cb}|$   {\rm vs.} $\alpha$
{\rm and} $\beta$ {\rm for} $\delta_b=0$\hfill& \cr
\+&\hfill\hfill& \cr
}$$
\epsfysize=7cm
\centerline{\epsffile[72 180 540 620]{/mercure/hamzaoui/pvcbc1.eps}}
\vfill\vfill\eject
$$\vbox{\settabs
\+&ddddddddddddddddddddddddddddddddddddddddddddddddddddddddddddddd&\cr
\+&{\rm Fig. 5}\hfill& \cr
\+&\phantom{ppp}$|V_{cb}|$  {\rm vs.} $\alpha$ {\rm and} $\beta$
{\rm for} $\delta_x={\pi\over 2}$ {\rm and } $\delta_b=0$&\cr
\+\hfill\hfill&\cr
}$$
\epsfysize=7cm
\centerline{\epsffile[72 180 540 620]{/mercure/hamzaoui/pvcbc2.eps}}
\vskip 0.5cm
$$\vbox{\settabs
\+&ddddddddddddddddddddddddddddddddddddddddddddddddddddddddddddddd&\cr
\+&{\rm Fig. 6}\hfill& \cr
\+&\phantom{ppp} {\rm A lower bound on} $|V_{ub}|$ {\rm vs.}
$\alpha$ {\rm and } $\beta$ {\rm for} $\delta_b=0$\hfill&\cr
\+&\hfill\hfill& \cr
}$$
\epsfysize=7cm
\centerline{\epsffile[72 180 540 620]{/mercure/hamzaoui/pvubc1.eps}}
\vfill\vfill\eject
$$\vbox{\settabs
\+&ddddddddddddddddddddddddddddddddddddddddddddddddddddddddddddddd&\cr
\+&{\rm Fig. 7}\hfill& \cr
\+&\phantom{ppp}$|V_{ub}|$ {\rm vs.} $\alpha$ {\rm and } $\beta$ {\rm for}
$\delta_x={\pi\over 2}$ {\rm and } $\delta_b=0$ \hfill&\cr
\+&\hfill\hfill& \cr
}$$
\epsfysize=7cm
\centerline{\epsffile[72 180 540 620]{/mercure/hamzaoui/pvubc2.eps}}
\vskip 0.5cm
$$\vbox{\settabs
\+&ddddddddddddddddddddddddddddddddddddddddddddddddddddddddddddddd&\cr
\+&{\rm Fig. 8}\hfill& \cr
\+&\phantom{ppp} $|V_{us}|$ {\rm vs.} $\alpha$ {\rm and } $\beta$
{\rm for} $\delta_x={\pi\over 2}$ {\rm and } $\delta_b=0$ \hfill& \cr
\+&\hfill\hfill& \cr
}$$
\epsfysize=7cm
\centerline{\epsffile[72 180 540 620]{/mercure/hamzaoui/pvusc1.eps}}
\vfill\vfill\eject
}
{\baselineskip=6pt\lineskiplimit=6pt\lineskip=6pt
{\bf References}\bigskip
$$\vbox{\settabs
\+ddddd& dddddddddddddddddddddddddddddddddddddddddddddddddddddddddddddd&\cr
\+[1]\hfill& Particle Data Group, L. Montanet et. al., Phys. Rev. {\bf
D50},\hfill\cr
\+\hfill& 1195 (1994). \hfill\cr
\+\hfill&\hfill \cr
\+[2]\hfill& Particle Data Group, L. Montanet et. al., Phys. Rev. {\bf
D50},\hfill\cr
\+\hfill& 1315 (1994). \hfill\cr
\+\hfill&\hfill \cr
\+[3]\hfill& H. Fritzsch, Phys. Lett. {\bf 73B} 317 (1977),\hfill\cr
\+\hfill& Nucl. Phys. {\bf B155}, 189 (1979); L. F. Li,
Phys. Lett. {\bf 84B}, 461\hfill\cr
\+\hfill& (1979); B. Stech, Phys. Lett. {\bf 130 B}, 189 (1983);\hfill \cr
\+\hfill& M. Gronau, R. Johnson and J. Schechter,
Phys. Rev. Lett. {\bf 54}\hfill \cr
\+\hfill& 2176 (1985); G. C. Branco and L. Lavoura,
Phys. Rev. {\bf D44}, 582\hfill \cr
\+\hfill& (1991); S. Dimopoulos,
L.J. Hall and S. Raby, Phys. Rev. Lett. {\b 68},\hfill \cr
\+\hfill& 1984 (1992); Phys. Rev. {\bf D45}, 4192 (1992);\hfill \cr
\+\hfill&\hfill \cr
\+\hfill& X.G. He and W. S. Hou, Phys. Rev. {\bf D41}, 1517 (1990);\hfill \cr
\+\hfill& R. E. Shrock, Phys. Rev. {\bf D45}, 10 (1992). \hfill\cr
\+\hfill&\hfill \cr
\+[4]\hfill& H. Fritzsch, Phys. Lett. {\bf 73B} 317 (1977),
Nucl. Phys. {\bf B155},\hfill\cr
\+\hfill& 189 (1979).\hfill\cr
\+\hfill&\hfill \cr
\+[5]\hfill& C. H. Albright, B. A. Lindholm and C. Jarlskog, \hfill \cr
\+\hfill& Phys. Rev. {\bf D38}, 872 (1988). \hfill\cr
\+\hfill&\hfill \cr
\+[6]\hfill& F. Abe et al., Phys. Rev. {\bf D50}, 2966 (1994). \hfill\cr
\+\hfill&\hfill \cr
\+[7]\hfill& Suraj N. Gupta and James M. Johnson, Phys. Rev. {\bf D44}
, 2110 (1991);\hfill\cr
\+\hfill& S. Rajpoot, Mod. Phys. Lett. {\bf A7}, 309 (1992);\hfill\cr
\+\hfill&Dongsheng Du and Zhi-zhong Xing, Phys. Rev. {\bf D48},\hfill \cr
\+\hfill& 2349 (1993).\hfill\cr
\+[8]\hfill& M. Kobayashi and T. Maskawa, Prog. Theor. Phys. {\bf 49},
652 (1973). \hfill\cr
}$$
}
\end